\documentclass[twoside,english]{dis09}
\usepackage[T1]{fontenc}
\usepackage[latin1]{inputenc}
\usepackage{verbatim}
\usepackage{floatflt}
\usepackage[dvips]{graphicx}

\makeatletter

\usepackage{wrapfig}
\usepackage{rotating}
\usepackage{array}

\usepackage{babel}
\makeatother
\begin{document}

\title{Progress in CTEQ/TEA global QCD analysis}

\author{P. M. Nadolsky,$^{a}$ J.~Huston,$^{b}$ H.-L.~Lai,$^{c}$ J. Pumplin,$^{b}$
and C.-P.~Yuan$^{b}$ \vspace{0.3cm}
\\
$^{a}$Department of Physics, Southern Methodist University, Dallas,
TX 75275, USA\vspace{0.1cm}
\\
$^{b}$ Department of Physics \& Astronomy, Michigan State University,
E. Lansing, MI 48824, USA\vspace{0.1cm}
\\
 $^{c}$ Taipei Municipal University of Education, Taipei, Taiwan }

\maketitle
\begin{abstract}
We overview progress in the development of general-purpose CTEQ PDFs.
The preprint is based on four talks presented by H.-L.~Lai and P.~Nadolsky
at the 17th International Workshop on Deep Inelastic Scattering and
Related Subjects (DIS 2009).
\\
\begin{center}
August 30, 2009
\end{center}
\end{abstract}

Interpretation of data from high-energy colliders, such as the Tevatron
at Fermilab and the LHC at CERN, relies on the knowledge of parton
distribution functions (PDFs) describing momentum distributions of
quarks and gluons in the hadron. The most comprehensive method for
determination of PDFs is based on a {}``global analysis'' of many
kinds of experiments, whose results are tied together by the theory
of Quantum Chromodynamics (QCD). Historically, two families of general-purpose
unpolarized PDFs have been in wide use, provided by the CTEQ \cite{Nadolsky:2008zw}
and MRST/MSTW \cite{Martin:2009iq} global analyses.%
\footnote{A third independent global analysis by Neural Network PDF Collaboration
\cite{Ball:2008by} is expected to be released in the near future. %
} Members of CTEQ (Coordinated Theoretical-Experimental project on
QCD) concurrently explore several kinds of PDFs \cite{Keppel:DIS09SF,Schienbein:2009kk,deFlorian:2009vb}.
Among these efforts, our group established and led for a long time
by Wu-Ki Tung (which we will refer to as {}``Tung et al.'', or {}``TEA'',
group) traditionally focuses on determination of general-purpose unpolarized
proton PDFs and their uncertainties. The TEA group has presented four
talks about its recent work at the DIS'2009 workshop in April 2009,
with the slides available from the workshop website \cite{Nadolsky:DIS09SF,Nadolsky:DIS09HQ,Lai:DIS09PT,Lai:DIS09MODLO}.
The purpose of this contribution is to summarize those presentations.
The main text will refer to the slides from the talks; it is essential
to have them open when reading the paper. 

\textbf{CT09 set of parton distributions.} In 2008, CTEQ/TEA group
released a CTEQ6.6M best-fit PDF set \cite{Nadolsky:2008zw}, together
with 44 supplementary PDF parametrizations needed for computation
of PDF uncertainties. Our next set of PDFs to be released, tentatively
designated {}``CT09'', will include a number of new features not
available in CTEQ6.6. It will incorporate new experimental data that
became recently available from the Tevatron Run-2 and address a number
of physics issues impacting the behavior of the PDFs. 

\textbf{\textit{\emph{Tevatron jet production.}}} The gluon distribution
$g(x,\mu)$ remains among the least constrained parton distribution
functions (PDFs), despite its important role in high-energy collider
physics. Its behavior is most uncertain at momentum fractions $x$
of order 0.1 or above, where the gluon PDF is constrained in the context
of a global QCD analysis largely by the data on inclusive single-jet
production at the Tevatron collider, $p\bar{p}\rightarrow\mbox{jet}+X$.
Considerable freedom in the parametrization of $g(x,\mu)$ at large
$x$ values, still allowed by the global analysis, affects predictions
for the Tevatron and LHC processes sensitive to gluon scattering,
including production of high-$p_{T}$ hadronic jets, $t\bar{t}$ pairs,
and Higgs bosons.

Recently, the CDF and D0 collaborations at the Tevatron Run-2 published
new measurements of inclusive single-jet production cross sections
\cite{Aaltonen:2008eq,Abulencia:2007ez,:2008hua}. They have smaller
statistical errors and better understood systematical uncertainties
in comparison to similar Run-1 measurements \cite{Affolder:2001fa,Abbott:2000kp}
available to CTEQ6.6. The impact of these Run-2 data on the gluon
distribution were investigated as a part of the CT09 fit \cite{Pumplin:2009nk}
and reviewed at the DIS'2009 workshop \cite{Nadolsky:DIS09SF}. We
focused, in particular, on the suggestion \cite{Martin:2009iq,:2008hua}
that the Run-2 data cause significant changes in the gluon PDF as
compared to CTEQ6.6. If, for example, the Run-2 jet data prefer a
smaller magnitude of $g(x,\mu)$ at large $x$, as suggested by the
MSTW'2008 study, it would reverse the trend followed by the CTEQ5
and CTEQ6 PDF series that tended to have an enhanced gluon distribution
in order to better accommodate the Run-1 jet cross sections. 

In contrast to those suggestions, the CT09 analysis does not confirm
that the Run-2 data necessitates suppression of $g(x,\mu)$ at large
$x$. The main role of these data is to impose significant constraints
on $g(x,\mu)$ and to reduce uncertainty in $g(x,\mu)$ at $x>0.1$
despite relaxation of subjective assumptions about the gluon parametrization
form made in the earlier global fits. 

The CT09 best fit achieves excellent agreement with the Run-2 data
\cite[slides 5a, 5b]{Nadolsky:DIS09SF}, while also preserving tolerable
agreement with the Run-1 data sets. The CT09 gluon PDF agrees with
CTEQ6.6 within the uncertainty bands \cite[slides 6 and 7]{Nadolsky:DIS09SF}.
The most significant difference between the CT09 and CTEQ6.6 $g(x,\mu)$
is observed at $\mu=Q<5$ GeV and $x\approx0.3,$ where CT09 $g(x,\mu)$
is \emph{enhanced} (rather than decreased) in comparison to CTEQ6.6
\cite[slide 7, top left figure]{Nadolsky:DIS09SF}. No significant
differences between CT09 and CTEQ6.6 are observed at other $x$ values
or larger $\mu.$ This can be contrasted to the behavior of MSTW'08
NLO PDFs, also fitted to the Run-2 jet data and shown in the right
figure of the same slide. While also enhanced above CTEQ6.6M at $x<0.3,$
the MSTW'08 NLO gluon PDF comes out to be systematically smaller than
CTEQ6.6M at larger $x$. This discrepancy in the conclusions about
the $x>0.3$ region -- suppression of MSTW'08 NLO $g(x,\mu)$ with
respect to CTEQ6.6$,$ and no discernible difference between CT09
and CTEQ6.6 -- can be more a reflection of different choices made
by the two groups (notably, in the selection of the functional form
for $g(x,\mu)$) than of a genuine constraint by the Run-2 data (which
become increasingly uncertain toward larger $x$). 

\emph{Theoretical uncertainty.} In light of the smallness of Run-2
statistical errors, uncertainty in $g(x,\mu)$ is now dominated by
systematic factors, each of which needs to be explored in turn. To
start, one must confirm that theoretical predictions for NLO jet production
cross sections available from several groups \cite{Ellis:1992en,Giele:1993dj,Nagy:2001fj,Nagy:2003tz}
agree within the desired accuracy of the analysis. We found \cite{Pumplin:2009nk}
that differences between the EKS and FastNLO NLO calculations \cite{Ellis:1992en,Kluge:2006xs}
employed in the global analysis are smaller than, or comparable to,
other systematic effects. At the same time, theoretical uncertainties
due to factorization scales, hadronization, and underlying event can
modify the jet cross sections by 10-20\% \cite{Olness:2009qd}. Unimportant
at present, this theoretical uncertainty will need to be accounted
for in the future and eventually reduced by calculating jet cross
sections at NNLO.

\emph{Agreement between single-jet production experiments; choice
of PDF parametrization.} The possibility of disagreement between the
Run-1 and Run-2 inclusive jet data was repeatedly raised as an explanation
for real or superficial differences between the pre- and post-Run-2
solutions for the gluon PDF. The best-fit gluon distribution obtained
in a simultaneous fit to the the Run-1 and Run-2 jet data sets agrees
with both; but, if fitted on their own, the Run-1 and Run-2 data prefer
somewhat different shapes of $g(x,\mu)$. The degree of (dis)agreement
depends on the functional form parametrizing $g(x,\mu_{0})$ at the
initial scale $\mu_{0}$. It can be exacerbated if the parametrization
of $g(x,\mu)$ is insufficiently flexible. 

To reduce the parametrization bias, CT09 employs a more flexible form
for $g(x,\mu_{0})$ than CTEQ6.6. This parametrization, designated
as {}``par1'' in \cite[slide 15]{Nadolsky:DIS09SF}, includes five
free parameters, vs. three parameters in CTEQ6.6. It leads to the
best agreement with the experimental data sets --- and it allowed
us to systematically explore agreement between the jet experiments
using a {}``$\chi^{2}$ reweighting method'' proposed in \cite{Collins:2001es}
and a new {}``data set diagonalization'' method described below. 

This investigation shows that the Run-1 and Run-2 jet experiments
are consistent, for the most part, with one another, with theory,
and with non-jet experiments. Yet, peculiarities of presently unknown
origin in the fit to the Run-1 data sets were also detected, which
may explain their preference for a somewhat different gluon PDF. A
persistently large $\chi^{2}/\mbox{d.o.f.}$ is observed in the case
of the CDF Run-1 data regardless of the PDF parametrization used,
reflecting excessive irregular scatter of these data. For the D0  Run-1
jet data, the $\chi^{2}$ value  varies excessively when very similar
CTEQ parametrizations are compared. These effects may reflect underestimated
uncorrelated errors present in the two measurements and,
in the case of D0 Run-1, peculiar dependence on the cross section
normalization and other correlated factors. Altogether, they may indicate
insufficient control of systematic effects in Run-1 jet production.

In contrast to {}``par1'', less flexible gluon parametrizations
tend to exaggerate tensions between the jet experiments, or between
the jet and non-jet experiments. With such parametrizations, a good
fit to all data sets required for the consistency study could not be attained.
Consider, for example, a $\chi^{2}$ reweighting scan exploring the
agreement between the jet and non-jet data. The scan multiplies the
jet data contribution $\chi_{jet}^{2}$ to the global $\chi^{2}$
by a weighting factor $w_{jet}$: \[
\chi^{2}=w_{jet}\,\chi_{jet}^{2}+\chi_{non-jet}^{2}.\]
Parametric dependence of $\chi_{jet}^{2}$ and $\Delta\chi_{non-jet}^{2}\equiv\chi_{non-jet}^{2}-\min(\chi_{non-jet}^{2})$
on $w_{jet}$ is shown in the figure on slide 15 of \cite{Nadolsky:DIS09SF}.
The CT09 parametrization, {}``par1'', is compared to three-parameter
(less flexible) parametrizations {}``par2'' and {}``par3'' defined
on the same slide. Variations in $\chi^{2}$ by 50-100 are deemed
statistically significant in this comparison, in accordance with the
CTEQ tolerance criterion \cite{Pumplin:2002vw}. Without the jet data
included ($w_{jet}=0$), the three parametrizations describe the non-jet
data sets equally well ($\Delta\chi_{non-jet}^{2}\approx0$ for all
three of them). When $w_{jet}$ is increased to 1 (the jet and non-jet
data are included on the same footing) or 10 (agreement with the jet
data is strongly emphasized over the non-jet data), the {}``par1''
form allows to significantly reduce $\chi_{jet}^{2}$, while only
modestly increasing $\chi_{non-jet}^{2}$ ($\Delta\chi_{non-jet}^{2}<50$).
This is less possible with the {}``par2'' and {}``par3'' forms:
the reduction in $\chi_{jet}^{2}$ happens only at the expense of
a significant increase in $\chi_{non-jet}^{2}$, especially for {}``par3''. 

HERA fits to DIS data commonly employ the {}``par3'' form to parametrize
the gluon distribution \cite{Chekanov:2005nn,H1:2009kv,Radescu:2009zz}.
The total number of \emph{all} free PDF parameters allowed in these
fits cannot be larger than about 10 due to the limited constraining
power of the DIS data fitted on its own. 
The $\chi^{2}$ reweighting scan shows that
the {}``par3'' form is flexible enough to describe the DIS data
on their own, but not to reconcile the DIS and Tevatron jet production
constraints in a simultaneous global fit. In the latter case, a more
flexible PDF parametrization such as {}``par1'' can be introduced
to provide a more realistic estimate of the PDF uncertainty.%
\footnote{In the NNPDF method \cite{Ball:2008by}, which operates with ultra-flexible
PDF parametrizations, the primary effect of the jet data is to reduce
the large uncertainty allowed for $g(x,\mu)$ by DIS experiments. %
}

\textbf{Data set diagonalization.} The analysis of data sometimes
requires fitting many free theoretical parameters to a large number
of data points. Questions naturally arise about the compatibility
of specific subsets of the data, such as those from a particular experiment
or those based on a particular technique, with the rest of the data.
Questions also arise about which theory parameters are determined
by specific subsets of the data. In Ref.~\cite{Pumplin:2009nm},
an extension of the Hessian method for uncertainty analysis \cite{Pumplin:2000vx,Pumplin:2001ct}
dubbed {}``data set diagonalization'' (DSD) was developed in order
to examine both kinds of questions. 

The DSD procedure identifies the directions in the PDF parameter space
along which a given subset $\mathbf{S}$ of data provides significant
constraints in a global fit. The procedure involves {}``a secondary
diagonalization'' of $\chi^{2}$ to obtain a new set of fitting parameters
$\{ z_{i}\}$ that are linear combinations of the original ones. In
the $\{ z_{i}\}$ representation, the data set $\mathbf{S}$ from
a given experiment (or another subset of the full data) and its complement
$\mathbf{\overline{S}}$ take the form of independent measurements,
within the scope of the quadratic approximation applied to $\chi^{2}$.
The degree of consistency between $\mathbf{S}$ and $\mathbf{\overline{S}}$
can thus be examined straightforwardly by comparing two independent
constraints on each parameter $z_{i}$ considered. 

When applied to practical fits \cite{Pumplin:2009nm}, the DSD method
uncovered and quantified the degree of tension between the two Tevatron
Run-2 inclusive jet experiments, and between one of those experiments
and the non-jet data, which was difficult to detect using the older
$\chi^{2}$ rescaling method. The DSD method can also identify which
features of the fit are controlled by particular experiments or other
subsets of the full data. As an example of this, the jet experiments
were shown to be the principal source of information on the gluon
distribution, by showing that the eigenvector directions dominating
the uncertainty of the gluon distribution are the same directions
that are constrained by the jet data (cf. Fig. 4 in \cite{Pumplin:2009nm}).
More generally, the DSD method is suitable for a systematic study
of consistency between experiments included in the full global fit
(again within the scope of the Gaussian approximation) \cite{Pumplin:2009sc}.

\textbf{Correlation analysis of PDF uncertainties.} In many practical
applications, it may be necessary to know if the PDF uncertainty of
one quantity, $X,$ is related in any way to the PDF uncertainty of
another quantity, $Y.$ For example, one may need to establish if
the PDF dependence of a specific cross section ($X=\sigma$) is driven
by a specific PDF parameter ($Y=z_{i}$), constrained in turn by the
measurement of another cross section, $Z=\sigma'.$ These (often entangled)
relations between the PDF dependence of different quantities can be
elucidated by yet another technique extending the Hessian method,
the PDF correlation analysis developed in Refs.~\cite{Nadolsky:2008zw,Pumplin:2001ct,Nadolsky:2001yg}.
The advantage of the correlation method as compared to the DSD or
Lagrange multiplier methods is that publicly available error PDF sets
are sufficient for carrying it out; it does not require specialized
statistical tools. 

In this approach \cite[slides 22-24]{Nadolsky:DIS09SF}, one computes
a cosine of correlation angle $\varphi$ for $X$ and $Y$, \[
\cos\varphi=\frac{1}{4\Delta X\,\Delta Y}\sum_{i=1}^{N}\left(X_{i}^{(+)}-X_{i}^{(-)}\right)\left(Y_{i}^{(+)}-Y_{i}^{(-)}\right),\]
where $\Delta A=\sqrt{\sum_{i=1}^{N}\left(A_{i}^{(+)}-A_{i}^{(-)}\right)^{2}}\Bigr/2$
is the usual PDF error for $A=X$ or $Y,$ and $A_{i}^{(\pm)}$ ($i=1,...N)$
are the values of $A$ evaluated for $2N=44$ error PDF sets. One
is most interested in situations when $\cos\varphi$ takes a value
close to $+1$ or $-1$. In these cases, the PDF dependence of $X$
and $Y$ is strongly correlated or anticorrelated, respectively, so
that an accurate measurement of $X$ would strongly reduce the PDF
uncertainty of $Y$.

\begin{wrapfigure}{r}{0.5\columnwidth}

\includegraphics[width=0.5\textwidth]{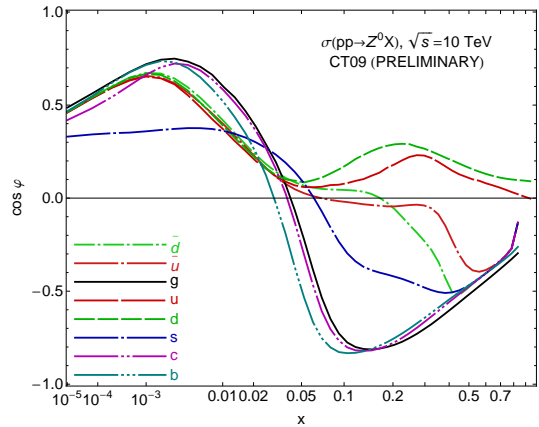}

\caption{Correlation between the NLO total cross section for $Z$
production at the LHC, $pp\rightarrow ZX$ at $\sqrt{s}=10$ TeV,
and PDFs $f_{a}(x,Q)$ at $Q=85$ GeV.\label{fig:zlhc10}}

\end{wrapfigure}

It is often instructive to calculate $\cos\varphi$ between a scattering
cross section $\sigma$ and a PDF $f_{a}(x,Q)$ at given $x$ and
$Q.$ One can identify the PDFs and the $x$ regions contributing
the bulk of the PDF error to $\sigma$ by plotting $\cos\varphi$
vs. $x$ and looking for the $x$ regions where $|\cos\varphi|$ is
large (say, above 0.7). Figure~\ref{fig:zlhc10} shows such plot
for the total cross section of $Z$ boson production at the LHC at
$\sqrt{s}=10$ TeV, while slide 24 in \cite{Nadolsky:DIS09SF} shows
the $Z$ cross section at four chosen values of rapidity $y$ of the
$Z$ boson. Each figure contains at least one curve with one or two
$x$ regions where $|\cos\varphi|$ is large. These $x$ regions and
PDF flavors drive the PDF uncertainty; the other regions are less
important. 

An interesting feature to notice is that the correlations
for the total $Z$ cross section are not reproduced exactly in either
bin of the rapidity of the $Z$ boson. The PDF uncertainty of the
total cross section in Fig.~\ref{fig:zlhc10} is correlated mostly
with the least known $c,$ $b,$ and $g$ PDFs at $x=10^{-3}-10^{-2}\sim M_{Z}/\sqrt{s},$
and not with the $u,$ $d$ quark and anti-quark PDFs that contribute
the bulk of the cross section, but are constrained much better \cite{Nadolsky:2008zw}.
The $Z$ cross section is also anti-correlated with the $c,$ $b,$
$g$ PDFs at $x\sim0.1,$ as a consequence of the momentum sum rule.
A similar correlation with the gluon and heavy-quark PDFs is observed
at central rapidities ($y=0.05$ and $1.05$ in slide 24), however,
these bins also show enhanced correlation with the strangeness distribution,
which is not present in the total cross section. At large rapidity
values of $y=2.05$ $y=3.85,$ a strong correlation develops with
the $u$ and $d$ (anti)quark PDFs, which overtakes the correlation
with the gluon and heavy-quark distributions. As $y$ increases, the
position of the largest correlations shifts from $x=10^{-2}$ at $y=0.05$
to $x=5\cdot10^{-4}$ at $y=3.85.$ The point of this exercise is
to show that the PDF uncertainty of the total cross section is only
an approximate predictor of the PDF dependence in the given process.
More detailed studies may be needed when pronounced kinematical dependence
of the PDF uncertainty is anticipated.

\textbf{Intermediate-mass scheme for heavy quarks.} A few-percent
accuracy expected from the modern (N)NLO PDFs requires, among other
things, to correctly implement heavy-quark mass contributions in cross
sections in the whole energy range considered in the global analysis
\cite{Nadolsky:2008zw,Martin:2009iq,Tung:2006tb,Thorne:2008xf}. This
implementation is realized most systematically in the general-mass
(GM) factorization scheme \cite{Collins:1998rz,Aivazis:1993pi}, a
recent implementation \cite{Tung:2006tb} of which is used in the
CTEQ6.6 and CT09 fits. A good fraction of input precision data from
DIS and fixed-target experiments included in the global analysis are
at energy scales comparable to, or not too far above, the charm and
bottom masses. The PDF parametrizations are thus sensitive to the
more precise treatment of mass effects in the GM scheme. In turn,
predictions for the Tevatron and LHC cross sections, including the
benchmark $W$ and $Z$ cross sections, also depend on the treatment
of heavy-quark masses at low energies. 

Although the GM scheme is clearly superior in its consistency to the
zero-mass (ZM) scheme, it is considerably more complicated, and thus,
not as widespread as the ZM scheme. On the other hand, since the most
important heavy-quark mass effects reflect energy-momentum conservation
in production of heavy quarks near their mass threshold, key features
of the GM scheme might be potentially reproduced by implementing full
mass dependence in the kinematical part of heavy-quark cross sections,
while still assuming simplified zero-mass expressions for their dynamics
described by the matrix elements \cite{Thorne:2008xf}. This hybrid
approach combining simple ZM dynamics with full GM kinematics was
worked out in detail at NLO \cite{Nadolsky:DIS09HQ,Nadolsky:2009ge}.
The resulting effective formalism, which we proposed to call an {}``intermediate-mass
(IM) calculational scheme'', can be viewed either as an \emph{}improved
zero-mass formulation with general-mass kinematics of final states,
or a simplified general-mass formulation \emph{}with zero-mass hard
matrix elements. It can be applied to reproduce the essential heavy-quark
mass dependence by a modest modification of the zero-mass calculation. 

The exact implementation of the intermediate-mass approach is not
unique, but depends on the choice of a rescaling variable $\zeta$
\cite[slides 5-6]{Nadolsky:DIS09HQ}. For one specific choice of $\zeta$
(corresponding to the parameter $\lambda=0.15$), all essential features
of the general-mass scheme are
reproduced both in terms of the resulting PDFs and in terms of typical
physics predictions at the Tevatron and the LHC \cite[slides 7-10]{Nadolsky:DIS09HQ}. These findings show
that the IM scheme indeed brings the existing NLO analyses based on
zero-mass hard matrix elements closer to the general-mass formulation.
Nonetheless, dependence of the IM predictions on the form of the effective
rescaling variable underlines the phenomenological nature of this
approach. Although this dependence in principle also arises in the
general-mass formalism, it is less pronounced than in the phenomenological
IM formulation \cite[slide 10]{Nadolsky:DIS09HQ}. Thus, the additional
source of theoretical uncertainty due to the choice of the rescaling
variable hardly affects what we know about the GM formalism -- except
that, perhaps, it should be added to the other sources of theoretical
errors, such as scale dependence, when assessing the uncertainty of
the GM theoretical results.

\textbf{New data on vector boson production.} The CT09 global analysis
will include new data on heavy vector boson ($W,$ $Z$) production
produced in the Tevatron Run-2. This includes, first and foremost,
the CDF measurements of charged lepton asymmetry in $W$ boson production
\cite{Acosta:2005ud} and CDF $Z$ boson rapidity distributions \cite{Aaltonen:2009pc}.
At this stage, we do not include the D0 Run-2 data on $W$ boson charge
asymmetry \cite{Abazov:2008qv}, since they cannot be well described
by the global analysis. Agreement with these data can be only improved
at the cost of having a much worse fit to the Run-1 $W$-lepton asymmetry
data and non-Tevatron experiments. With the D0 Run-2 data in the fit,
the ratio of down-valence to up-valence parton distribution functions
in the large $x$ region becomes too large as compared to the CTEQ6.6
uncertainty band. On the other hand, the $W$-lepton asymmetry data
from CDF Run-2 agree well with the rest of the global analysis. CDF
also released a new measurement of the $W$ charge asymmetry as a
function of $W$ boson rapidity reconstructed from the 4-momentum
of the charged lepton \cite{Aaltonen:2009ta}. These data agree well
with the recent CTEQ6.X PDFs, and its primary role would be to reduce
the uncertainty in the relevant combinations of quark PDFs. For this,
theoretical uncertainty arising in the reconstruction of the $W$
boson rapidity distribution from the directly observed lepton distribution
will need to be better understood.

%
{}

\textbf{Combined PDF+$p_{T}$ fit.} Recently, we extended the conventional
global QCD analysis to include experimental data on transverse momentum
($P_{T}$) distributions in low-$Q$ Drell-Yan and $Z$ boson production
in hadron-hadron scattering \cite{Lai:DIS09PT}. The $P_{T}$ data
is described systematically by Collins-Soper-Sterman resummation \cite{Collins:1984kg},
which accounts for soft QCD radiation that affects multi-scale measurements.
Resummation introduces a phenomenological function $S_{NP}(b,Q)$
to describe nonperturbative QCD effects parametrized in impact parameter
($b$) space. Within the combined PDF+$P_{T}$ analysis, we are able
to pursue determination of the nonperturbative function $S_{NP}$
and the PDFs simultaneously. It is expected to give a better estimate
for quantities that rely on both, especially for $W$ boson mass measurement.
As one of preliminary results, we obtained a new estimate of $S_{NP}(b,Q)$
at the invariant mass $Q=M_{Z}$ relevant for $Z$ boson production
in the Tevatron Run-2. By parametrizing $S_{NP}(b,M_{Z})=gb^{2}$
and applying the Lagrange multiplier method, we found $g=2.49_{-0.39}^{+0.24}$,
which is slightly larger than the value of $g$ reported by D0 collaboration
\cite{:2007nt}. We are in the process of producing a specialized
set of error PDFs that will include additional PDF eigenvector sets
to describe the uncertainty in $S_{NP}(b,Q)$. 

\begin{floatingfigure}[o]{0.5\columnwidth}%
\includegraphics[width=0.5\columnwidth]{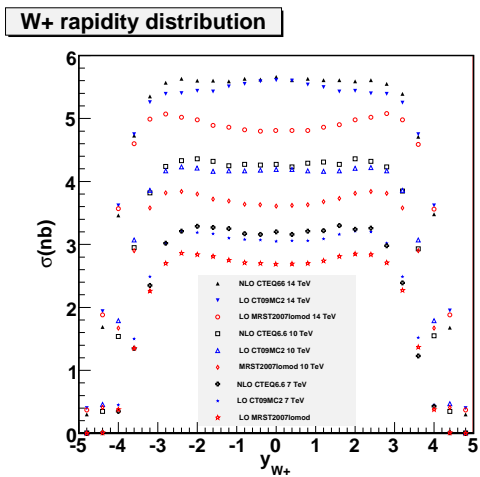}

\caption{The $pp\rightarrow W^{+}X$ cross section at $\sqrt{s}=10$ and $14$
TeV, from the NLO/CTEQ6.6 calculation and LO calculations based on
CTEQ and MSTW modified LO PDFs. \label{fig:ModLO}}\end{floatingfigure}%

\textbf{PDFs for leading order Monte-Carlo showering programs.} Just
as NLO PDFs are used with NLO calculations, it has been natural to
use LO PDFs with LO calculations, including the popular parton shower
Monte Carlo programs. Unfortunately, many collider matrix element
calculations realized at LO differ from NLO predictions not only in
magnitude, but also in shape, because of the impact of hard-scattering
corrections, but also because of the differences existing between
the LO and NLO PDFs. This problem has led to the introduction of PDFs
that are specially designed for leading-order Monte Carlo (LO MC)
programs. They attempt to lessen the differences between the LO and
NLO predictions by modifying the PDFs. Such LO MC PDFs have been produced
recently by CTEQ/TEA \cite{Lai:DIS09MODLO}. In our approach, pseudodata
for NLO \emph{theoretical} cross sections for several benchmark collider
processes have been included in the global fit along with the usual
experimental data. Specific pseudodata sets considered were for $W$,
$Z$, $t\bar{t}$, $b\bar{b}$, and Standard Model Higgs boson production
in $pp$ collisions at 14 TeV, generated at NLO using the CTEQ6.6
set of PDFs. The purpose of the pseudodata is to enhance the desired
NLO behavior in the LO MC PDFs. Two approaches were adopted: (1) the
momentum sum rule was kept intact, and the scales for the pseudodata
cross sections were varied within a limited range to achieve the best
fit to the shape of the relevant NLO kinematic distributions; or (2)
violations of the momentum sum rule were allowed, and the best LO
fit to the pseudodata cross sections was attempted, in terms of both
normalization and shape. An example of the latter approach is shown
in Figure \ref{fig:ModLO}, where the $W^{+}$ rapidity distribution
at the LHC is shown for center-of-mass energies of 14, 10 and 7 TeV.
The CTEQ LO MC PDF (CT09MC2) leads to better agreement with the fully
NLO prediction at 14 TeV than the LO prediction using the MRST2007lomod
PDF, both in the shape of the $W^{+}$ rapidity distribution and in
its normalization. Good agreement is also observed for the predictions
for the two lower center-of-mass energies for the LHC. A similar level
of agreement is achieved for the $W^{-}$ and $Z$ cross sections,
and better shapes and normalizations are obtained for the heavy-flavor
and Higgs cross sections as well. 

\textbf{PDF reweighting and FROOT.} In a typical calculation of the
PDF uncertainty, the user must compute the cross section of interest
for a large number ($N=30-1000$) of error PDFs. A multi-loop QCD
calculation for \emph{even one PDF set} can require substantial computer
resources. Straightforward repetition of this calculation for \emph{many
PDF sets} can quickly become intractable. 

This bottleneck can be eliminated by a general-purpose technique for
event reweighting in Monte-Carlo integration programs. The event reweighting
reduces the CPU time needed to compute all $N$ cross sections by
evaluating complicated multi-loop matrix elements only once for all
$N$ PDFs. The event reweighting also improves convergence of the
Monte-Carlo estimate for the PDF uncertainty \cite[slides 25-26]{Nadolsky:DIS09SF}.

The TEA group develops FROOT \cite{FROOT}, a publicly available library
that facilitates PDF reweighting and analysis of large numerical outputs
for many PDF sets in theoretical calculations written
in Fortran or C++. The current version of FROOT provides a simple interface
to write differential cross sections for $N$ PDF sets into CERN ROOT
trees, which can then be analyzed inside the ROOT program 
to evaluate PDF uncertainties,
PDF correlations, etc. Depending on the setup of the calculation,
this approach can substantially reduce (by a factor of 10 or more) the
requirements for CPU time and hard-disk storage requirements.
The FROOT interface is implemented in MCFM \cite{Campbell:2000bg,Campbell:2004ch,Campbell:2005bb}
and ResBos programs \cite{Balazs:1997xd,Landry:2002ix} computing
a variety of theoretical cross sections. In the future, FROOT will
be expanded to include additional features facilitating the analysis
of the PDF dependence of collider cross sections in CERN ROOT.

\subsection*{Acknowledgments}

This research was supported by the U.S.\ Department of Energy under
grant DE-FG02-04ER41299; U.S. National Science Foundation under grants
PHY-0354838, PHY-055545, and PHY-0757758; National Center for Theoretical
Sciences and National Science Council of Taiwan grant NSC-97-2112-M-133-001;
by LHC Theory Initiative Travel Fellowship awarded by the U.S. National
Science Foundation under grant PHY-0705862; and by Lightner-Sams Foundation.


\providecommand{\href}[2]{#2}\begingroup\raggedright
\endgroup

\end{document}